\documentclass[conference]{IEEEtran}
\IEEEoverridecommandlockouts
\usepackage{cite}
\usepackage{amsmath,amssymb,amsfonts}
\usepackage{cuted}
\usepackage{graphicx}
\usepackage{textcomp}
\usepackage{fancyhdr}
\usepackage{multirow}
\usepackage[table]{xcolor} 
\def\BibTeX{{\rm B\kern-.05em{\sc i\kern-.025em b}\kern-.08em
    T\kern-.1667em\lower.7ex\hbox{E}\kern-.125emX}}

\begin{document}

\title{Towards a Better Modelling of the Cell Formation Problem: An Overview of Decisions and a Critical analysis of Constraints and Objectives* 
\thanks{* This is an almost integral translation of a french language paper presented in a conference held in Algeria \cite{Boulif2010}. Despite being a little outdated, we think it contains some material that is still useful for the community.}\\}

\author{
\IEEEauthorblockN{BOULIF Menouar}
\IEEEauthorblockA{\textit{Department of Computer Science } \\
\textit{M'Hamed Bougara University of Boumerdes}\\
Independence Avenue, 35000,
Boumerdes, Algeria \\
boumen7@gmail.com}

\and
\IEEEauthorblockN{ATIF Karim}
\IEEEauthorblockA{\textit{Department of Computer Science} \\
\textit{Houari Boumediene University}
\textit{of Sciences and Technology}\\
El-Alia Babezzouar, 16111, Algiers, Algeria.\\
atif\_karim2001@yahoo.fr}
}

\maketitle

\begin{abstract}
Cell formation problem is among the first obstacles the designer of cellular production systems must overcome. This paper presents a critical analysis of the various criteria and constraints considered in the literature. The objective is to help any researcher who wants to model the problem by adopting a multi-criteria approach.
\end{abstract}


\section{Introduction}
Cellular Manufacturing Systems are an implementation of the Group Technology philosophy (TG). In the design of these systems, the problem of cell formation consists in decomposing the set of resources (machines in general) of a manufacturing workshop into production units called cells. Each cell will be associated with a family of products with similar characteristics. This decomposition is broadly speaking carried out with the aim of optimising the competitiveness of the plant.\\
The cellular organisation of the production plants makes it possible to improve the quality of the products because the responsibility for the quality of a given family lies on the team of operators assigned to the associated cell: each operator has the possibility of stopping the process of manufacturing if it detects any anomaly. On the other hand, the fact of making similar products in the same cell makes it possible to reduce the manufacturing time because the time for adjusting the machines is reduced due to the low recourse to changing tools. This has the consequence of drastically improving the efficiency of the production.\\
Starting with the work of Burbiddge \cite{Burbidge1963} and then McAuley \cite{McAuley1972}, research in the field took off in the 1990s. Current trends are moving towards the adoption of more pragmatic approaches (integrated, multi-criteria, dynamic) as they try to get more closer to reality.
This article aims to help such approaches for a good modelling of the problem which is a necessary condition for the good solving of this one.\\
The rest of this article is organised as follows: Section two describes the various decisions to be taken to resolve the problem. The next section gives an overview of the constraints considered in the literature. Section four provides an overview of the criteria as well as a critical study that paves the way for a multi-criteria approach. Finally, we present our conclusion.
 
\section{Description of the decisions}
In the literature, solving the composition problem means making one of the following decisions:

\begin{itemize}
     \item Decomposition of machines into cells, and products into families and an assignment of families to cells (simultaneously or sequentially) \cite{King1980, Garcia1986, Sofianopoulou1997, Chen1999}. If these three actions are carried out at the same time, we are talking about a mixed composition problem.
     \item Only partitioning the products into families \cite{Al-Sultan1997}.
     \item Only partitioning the machines into cells \cite{Atmani1995, Pierreval1998, Mungwattana2000, Zhao2000, Vin2003, Boulif2006, Boulif2008}.
\end{itemize}

In some so-called integrated approaches, one or more of the following additional decisions are associated with the composition problem:
     
\begin{itemize}
     \item Determination of the number of machines to produce the planned products \cite{Logendran1997, Su1998, Wicks1999, Mungwattana2000, Defersha2008}.
     \item Selection of opertation sequence to be adopted for each product \cite{Atmani1995, Defersha2008}.
     \item Assignment of operations to the available machines \cite{Atmani1995, Vin2003, Defersha2008}.
     \item Description of the products to be subcontracted \cite{Mansouri2002}.
     \item Machine Layout inside the cells \cite{Akturk1996}.
     \item Assignment of machines to sites on the workshop \cite{Irani1993}.
     \item Assignment of operators to cells \cite{Min1993}.
\end{itemize}

In so-called dynamic approaches, these decisions must be made in each period of the planning horizon.\\ 

 Considering all these decisions in a dynamic model, and then solving it efficiently, remains an open problem.
\section{Analysis of the constraints}
For a solution to the composition problem to be adopted, it must satisfy certain conditions or constraints. Most of the time, the constraints are either imposed by actual considerations or for simplification reasons. In the first case, we are obliged to take them into account, otherwise the implementation of the solution in the plant would be impossible. In the second, it is quite the opposite because, for the sake of simplification, we can end up with "laboratory" models that bear little relation to real circumstances and almost useless in practice. To these two types of constraints, we can add two others, namely, criteria constraints and consistency constraints.\\

 In what follows, we present the various constraints taken into account in the literature.\\
 
\subsection{Practical constraints}
\subsubsection{Size of product families}
The size of a family is the number of products that it contains. It is preferable to moderate this size in order to properly master the quality objective by balancing the burden of product responsibility between the different teams in the plant. Among the works having considered this kind of constraints, we find \cite{Moon1999}. The authors cite as a pretext the quest for simplicity of control and planning in the cell. Other authors, such as \cite{Taboun1998, Wicks1999}, impose a lower bound on any family of products to be created.\\

\subsubsection{Size of machine cells}
This constraint is by far the most considered in the literature. The size of a cell is understood to mean the number of machines that it can house. This number should be moderate (and therefore limited) for various reasons generously cited in the literature. For example, \cite{Vakharia1990} argues that the space available in the workshop can limit the number of machines to put in each cell. In addition, the cell size should not be too large at the risk of compromising the socio-professional environment in the cell or making visual control in it difficult. The authors of \cite{Rajagopalan1975} give other considerations, such as the possibility of neglecting the costs of intracellular handling and the consolidation of the gains of Group Technology (reduction of work-in-progress, machine adjustment times, etc.) because with large cell sizes, these gains may disappear. In \cite{Nagi1990}, the author claims the restriction is linked to the capacities of manipulating robots and intracellular buffers which can only supply a moderate number of machines.\\

Among those who opted for this type of constraints, we find \cite{Rajagopalan1975, Askin1990, Harhalakis1990, Nagi1990, Boctor1991, Min1993, Harhalakis1994, Liang1995, Logendran1995, Murthy1995, Sofianopoulou1997, Su1998, Chen1998, Chen1999, Zhao2000}.\\

\subsubsection{Number of cells}
It is preferable to have a moderate number of cells to facilitate the management of the plant. To do this, two bounds can be used: a lower bound and an upper one. Among the research works having opted for these constraints, we find \cite{Adil1996, Chen1999, Boulif2006}. Note the interference between this constraint and that of the cell size. Indeed, the use of a bound for the maximum number of machines in each cell, induces a constraint on the minimum number of cells, equal to the first integer greater than the number of machines, divided by the maximum number of machines per cell. It must then be checked whether this number does not exceed the upper limit of the number of cells, in which case the search space would not contain any feasible solution. Consequently, if the cell size constraint is considered, we can omit the constraint of the lower bound of the number of cells for the previous reasons as well as that of the upper bound because such a bound always exists (number of machines) and hence, not setting it allows the optimisation process to search for the best suitable number of cells (There will be further discussion of this issue in \ref{fixNbrCell}).\\

\subsubsection{Cohabitation constraints}
This class of constraints involves two types of constraints: cohabitation constraints and non-cohabitation constraints. In the first type, it is a matter of considering the situation where we want to put some machines close to each other. The reason is that, for example, they require energy sources with expensive or bulky installations \cite{Boulif2006}. For the second type, it is a question of an aim to distance certain machines from each other. For example, to distance high precision machines from machines that generate large vibrations. Among the authors including in their models this class of constraints we find \cite{Souilah1994, Chen1999, Boulif2006}.\\

\subsubsection{Capacity of machines}
This kind of constraints ensures that the load of the machines does not exceed its capacity. A necessary condition for the consideration of this constraint to be plausible is the consideration of a certain flexibility of the routing such as:
\begin{itemize}
	\item the consideration of several machines of the same type, 
	\item the possibility of carrying out an operation on different machines or, in a generalised form, 
	\item the possibility of making each product according to several operating ranges. 
\end{itemize}

	This constraint is taken into account in \cite{Askin1990, Nagi1990, Harhalakis1994, Logendran1994, Atmani1995, Logendran1995, Taboun1998, Wicks1999, Vin2003}.\\

\subsection{Simplification constraints}
It is wrong to believe that adding constraints to a problem is always synonymous to increased difficulty. The cause of this thought is perhaps the additional cost of calculation generated by feasibility checking of the solutions in the solving methods that are based on total or partial enumeration. This means that commercial problem-solving software for mathematical programs endeavours to evaluate the efficiency of their products by the sum of the number of variables and constraints that they can handle. It should not be forgotten that a constraint leads to a reduction in the search space, and that in certain cases \cite{Boulif2006}, it is possible to solve an NP-hard problem efficiently if constraints are added to it. Among the simplification constraints that we encountered, we present the following.\\

\subsubsection{Fixed value family size}
Instead of limiting the size of families, Al-Sulatan \cite{Al-Sultan1997} considers fixed sizes for the product families. The authors give no practical explanation for their restriction. Of course, it seems that the concern is to limit the liability of cell operators to a moderate number of products. However, in this case, the restriction by imposing an upper bound would have been sufficient.\\

\subsubsection{Predetermined number of cells}
\label{fixNbrCell}
Most of the published research (for example, \cite{Atmani1995, Akturk1996, Su1998, Taboun1998, Zhao2000, Mungwattana2000, Mansouri2002, Defersha2008}) assumes that the number of cells (and therefore, if applicable, that of families) is fixed beforehand. Some argue that this is due to practical considerations such as the recommendations of the decision maker who wants to fix this number to simplify the management of his workshop. Others \cite{Mungwattana2000} clearly state that this is a simplifying assumption because it makes it possible to reduce the search space significantly. It is obvious that the practical considerations put forward by the former can be achieved by determining two bounds for the number of cells: a lower bound and another upper one. Among the authors who strongly adhere to the clumsiness of the a priori fixing of the number of cells, we have \cite{Taboun1991, Liang1995, Sofianopoulou1997, Jaya1998}.\\

\subsection{Criterion-derived constraints}
It is possible to take certain criteria into account, not by including them in the objective, but by associating them with constraints. Some authors have opted for this technique in their multi-criteria approaches. We have identified the following applications for such an approach:\\

\subsubsection{Budget constraint}
Assuming that the budget is limited, several studies require that the cost associated with one (or more) given objective (s) does not exceed the allocated envelope. In \cite{Logendran1995}, an application of such an approach considers the sum of the cost of duplicating machines with that of subcontracting products.\\

\subsubsection{Similarity threshold}
Instead of considering the similarity between products as a criterion to be maximised, the authors in \cite{Taboun1991, Taboun1998} consider for each family of product a threshold not to be crossed for the associated solution to be accepted.\\

\subsubsection{load balancing}
To balance cell loads, the authors of \cite{Defersha2008} use the  total load average multiplied by a certain factor as a lower bound on any cell load in order to prevent cell underuse. If the multiplication factor, which is between 0 and 1, is close to unity, cell loads are forced to be close to the average.\\

\subsection{Coherence constraints}
In addition to the previous types of constraints, we can find in the literature constraints that we call coherence constraints. These are especially used in mathematical models (most often in mathematical programming approaches) and are added to preserve the consistency of the variables in the model. By way of illustration, if a variable indicates the assignment of a machine to a cell, then there should be no variable value indicating that the machine is assigned to several cells at the same time. The fact that this kind of constraints depends on the model and that it must have no influence on the non-formal statement of the problem (because the associated restrictions are clearly implied), we did not consider it useful to present the different associated cases.

\section{Analysis of the criteria} 
Our choice of the expression "optimising competitiveness" used earlier was not accidental. It was necessary to use an expression as flexible as possible to incorporate the plethora of objective functions present in the literature. Indeed, as with any real problem, there is not actually a single objective to be achieved, but several. To be convinced of this reality, let's take a look at the criteria to optimise that are proposed in the literature.\\

\subsection{Overview of criteria}
We were able to identify the following criteria on more than thirty published articles, in addition to another twenty studied in \cite{Mansouri2000}. Unlike of the latter which is limited to multi-criteria approaches, our overview also concerns the criteria associated with mono-objective approaches.\\

\subsubsection{Minimisation of special elements}
This criterion is calculated by counting the ones outside the diagonal blocks of the product-machine incidence matrix (PMIM). They are called special elements insofar as they designate operations associated with products produced outside their cells, or in a dual way, with machines performing tasks on external products. The ideal case of zero-special-elements is equivalent to completely independent cells. Among the researchers who opted for this criterion, we find \cite{Boctor1991, Murthy1995, Adil1996}.\\

\subsubsection{Minimisation of intercellular movements}
This criterion seeks to minimise the movements that the products make between the cells. This criterion has been proposed to address an insufficiency related to the criterion of special elements minimisation. Indeed, the latter does not give the appropriate number of movements when a product performs non-consecutive operations on a machine external to its cell. Among the papers having used this criterion, we have \cite{Venugopal1992a, Shafer1992, Dahel1993, Gupta1995, Logendran1995, Sureh1995, Verma1995, Sofianopoulou1997, Zhao2000, Goncalves2002, Mansouri2002, AlBadawi2005}.\\

\subsubsection{Minimisation of intracellular movements}
This criterion is considered by some authors in order to avoid movements of products in the opposite direction or to have to bypass machines inside the cells. Indeed, this kind of movements causes complications in the handling system and a lack of security (risk of collision). In general, this criterion is taken into account along with minimising intercellular movements, the two combined by a weighting that favours the latter, in an aggregating objective function. Among the papers offering such a combination, we have \cite{Logendran1991, Logendran1995, Gupta1995, Verma1995, Gupta1996, Zhao2000}.\\

\subsubsection{Cost of machine duplication (or investment)}
Duplicating machines makes it possible to avoid intercellular movements, by acquiring for the products requiring an external operation a new machine that can achieve it. The price to pay for this duplication is an additional cost of investment. The criterion in question serves precisely to minimise the cost of this investment. Among the works including this criterion, we have \cite{Askin1990, Shafer1992, Logendran1994, Sureh1995, Logendran1997, Su1998, Wicks1999, Mansouri2002}.\\

\subsubsection{Cost of product subcontracting}
Another alternative which makes it possible to avoid intercellular movements is to subcontract the products requiring external operations. This subcontracting has a cost and this criterion seeks to minimise it. Note that in general, this criterion is combined with that of duplicating machines in order to find a compromise. Among the approaches that have opted for this criterion, we find \cite{Wei1990, Shafer1992, Logendran1997, Taboun1998, Mansouri2002}.\\

\subsubsection{Minimisation of intercellular traffic}
This criterion is close to the criterion for minimising intercellular movements. However, instead of considering the flows between machines equally, it takes into account the volume of the transported products or their cost. Among the works using this criterion we have \cite{Askin1990, Harhalakis1990, Nagi1990, Venugopal1992a, Harhalakis1994, Atmani1995, Pierreval1998, Su1998, Taboun1998, Chen1999, Wicks1999}.\\

\subsubsection{Maximisation of the use (or minimisation of underuse) of the machines inside the cells}
This criterion seeks to intensify the use of the machines in each cell by the products of the associated family. The majority of works having opted for this criterion proceed by the maximisation of ones (or the minimisation of zeros) inside the diagonal blocks of the PMIM matrix. This criterion is, in general, combined in a weighted manner with the minimisation of ones outside the diagonal blocks. This combination provides a performance measure which was widely used in early work, known as 'Grouping Efficiency' or GE (for Grouping Efficiency then Grouping Efficacy) \cite{Chandrasekharan1989, Kumar1990}. It has subsequently known several variants which have been analysed in \cite{Sarker2001}. The criterion in question was chosen in the papers \cite{Chen1999, Mansouri2002}. On the other hand, among the papers which used one of the variants of the GE combination, we have \cite{Chen1995, Mukhopadhyay1995, Joines1997, Zolfaghari1997, Goncalves2002}.\\

\subsubsection{Cost of product manufacturing}
This criterion generally comes into play when considering the possibility of carrying out an operation on different machines (routing flexibility case), each with a different cost (or time) of manufacture. We then seek to assign the products to the machines available so as to minimise the overall cost of production while meeting capacity constraints. Among the papers using this criterion: \cite{Sankaran1993, Atmani1995, Ho1996, Rajamani1996}.\\

\subsubsection{Minimisation of intracellular load variations}
The machine workload is the aggregated manufacturing time of the products assigned to it. This criterion therefore seeks to balance the workloads of the machines so that the cell operators can have similar workloads, on the one hand. On the other, to ensure a certain regularity of the flows inside the cells in order to minimise the work in progress \cite{Venugopal1992a}. Among the authors who chose this criterion, we have \cite{Wei1990, Venugopal1992a,Venugopal1992b, Akturk1996, Gupta1996, Su1998, Zhao2000}.\\

\subsubsection{Minimisation of intercellular load imbalance}
This criterion applies to the loads of the cells which are calculated by the sum of the workloads of the machines which compose them. We want, therefore, by this criterion to have cells with similar loads because balancing these loads makes it possible to reduce the size of the buffer stocks of intermediate components \cite{Akturk1996} and to avoid several problems related to the workshop management \cite{Mansouri2002}. Among the papers using this criterion, we find \cite{Wei1990, Akturk1996, Lee1997, Pierreval1998, Su1998, Mansouri2002}.\\

\subsubsection{Intracellular avoidance cost}
When a product does not use all the machines of its cell, an avoidance cost is generated by the need to use special handling means, the extension of manufacturing times and possibly the increase in work in progress \cite{Verma1995}. It is therefore wise to minimise this cost to comply with the Group Technology philosophy. Among the works which have chosen this criterion, we have \cite{Verma1995, Akturk1996}. To take this criterion into account, we need information on the layout of the machines inside the cells.\\
While quantitative criteria have been adopted by the majority of researchers, some have tried to include qualitative criteria in their model. The distinctive feature of qualitative criteria is, perhaps, the difficulty of giving a clear definition which cannot be disputed. Among the criteria we consider as being part of this category, we have:\\

\subsubsection{Maximisation of similarities (or minimisation of dissimilarities) between products of the same family}
This criterion adheres to the spirit of the GT philosophy. To measure the similarity (resp. Dissimilarity) between two products, we generally use a function which calculates the similarities (resp. dissimilarities), between the associated row vectors, of the incidence matrix PMIM. The pioneer of this technique is McAuley \cite{McAuley1972} who used the Jaccard's Similarity Coefficient ($JSC$) defined by equation \ref{jsc}.

\begin{strip}
\begin{equation}
\resizebox{0.6\textwidth}{!} {$JSC =\frac{Number\ of\ pairs\ whose\ components\ are\ equal\ to\ one} {Number\ of\ pairs\ with\ at\ least\ one\ component\ equal\ to\ one}$}
\label{jsc}
\end{equation}
\end{strip}

Fourteen similarity coefficients are analysed in \cite{Anderberg1973}. Another survey is done in \cite{Yin2001}. Relatively recent approaches that have opted for similarity coefficients use more elaborate formulas including information concerning production volumes, manufacturing times and operation sequences. \\
Among the works which have chosen this criterion, we cite \cite{Jaya1998, Seifoddini1989, Gupta1990, Taboun1991, Vakharia1990, Liang1995, Al-Sultan1997, Vin2003, AlBadawi2005}. To take this criterion into account, we need to undertake decisions on the allocation of products to families.\\

\subsubsection{Minimisation of the reconfiguration cost}
\begin{figure*}
\centering\includegraphics[scale=0.26]{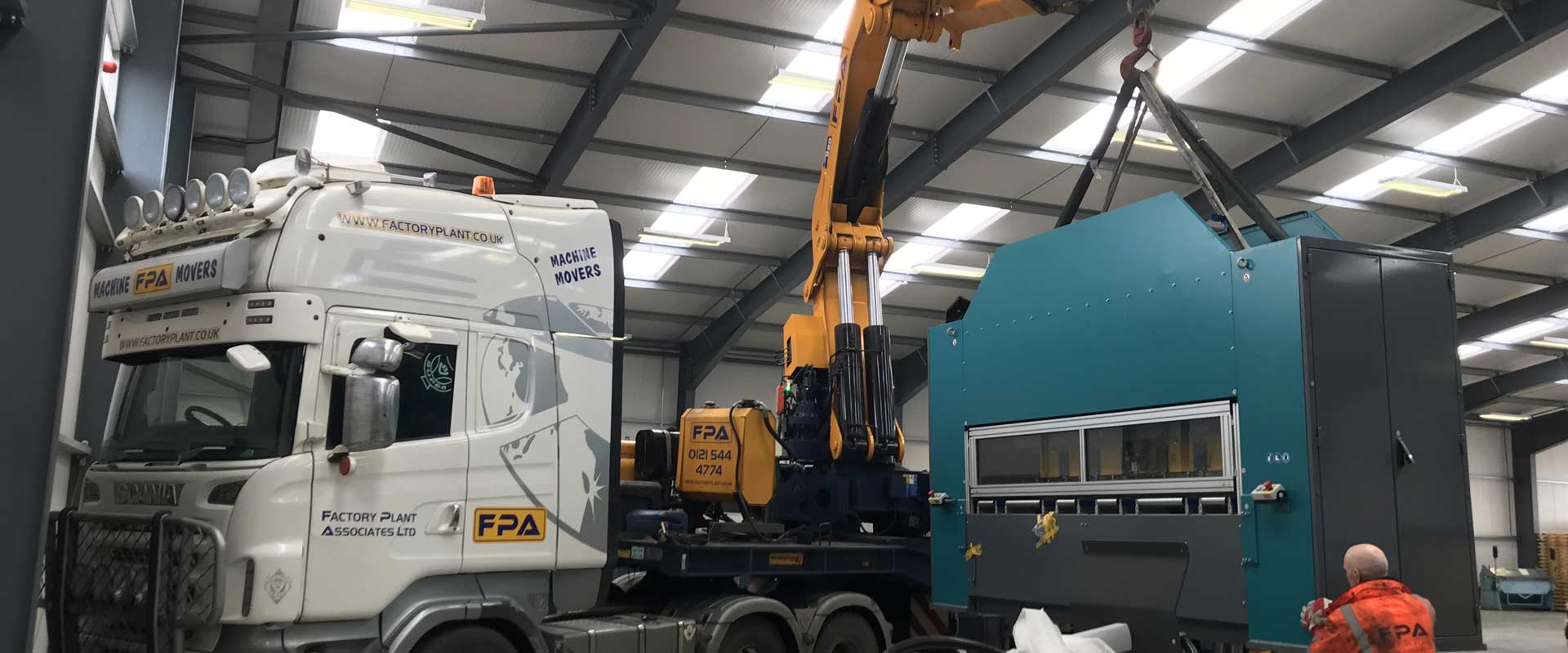} 
\caption{Moving of machines for cell reconfiguration may require special equipment (used by permission of the owner \cite{FPA_url})}
\label{machineMoving}
\end{figure*}

This criterion is exclusively considered by dynamic approaches. In this type of approaches, certain inputs (such as operation sequences and product demands) may change over time. Consequently, the planning horizon is subdivided into periods in which possibly different compositions would be implanted. This generates a reconfiguration cost which measures the double cost of the changes to be applied to the structure of the system and their effects (movement of machines (see figure \ref{machineMoving}), unproductive time). The authors in \cite{Chen1998, Wicks1999, Mungwattana2000, Defersha2008} calculate this cost in a classic way on the basis of moved machines. The authors of \cite{Boulif2008} after explaining the limits of the conventional method, calculate this cost indirectly based on a codification of the cell compositions.\\

\subsubsection{Maximisation of flexibility}
\label{maxFlex}
Although the term flexibility is a leitmotif in the literature dealing with cell formation, the following passage allows us to realise the difficulty of its definition: "Ten or fifteen years ago, quality was much like flexibility is today: vague and difficult to improve yet critical to competitiveness… Flexibility is only beginning to be explored… It means different things to different people" \cite{Upton1995}. We invite the reader, wishing to discover the different facets of this concept, to consult \cite{DeToni1998}. What interests us in this part is the way in which this criterion has been taken into account by the research works carried out in the field. We were able to identify two applications:\\
The first \cite{Liang1995} consists in measuring flexibility by the number of products that can be directly manufactured by the cellular system. Indeed, the authors propose an approach which consists in converting a job-shop system into a cellular system. Certain products of the original system, which could not be assigned to families during the optimisation process, are subcontracted. But instead of minimising the cost of subcontracting, the authors maximise the number of products held back. This allows us to draw the observation that the criterion thus defined is dual to the criterion of minimising the cost of subcontracting.\\
In the second application \cite{Vin2003}, the authors start from the fact that if a machine breaks down, it is useful to be able to reassign its workload without producing a great delay in production. To do this, the authors propose to define for each machine a threshold of use lower than the real capacity of the machine in question (a similar proposition is given in \cite{Vakharia1990}, but without alluding to its use to measure flexibility) . This usage threshold is not imposed as a strong constraint but by a penalty that measures the degree of violation of the threshold. Thus, each solution is assigned a degree of compliance with this aim, equal to the average of the penalties associated with all of the machines.\\
It is obvious that this process cannot be carried out with conventional approaches which do not consider alternative routings (the consideration of machine duplication is a particular case of routing flexibility) because without routing flexibility machine utilisation is constant.\\

\subsubsection{Maximising production efficiency} 
In \cite{Liang1995}, the authors assert that the efficiency of a production system is the result of the simplification of production flows, the reduction of handling tasks and manufacturing times. Consequently, they consider that the measurement of efficiency involves the evaluation of these three parameters. However, they say, the determination of these parameters in the cell composition step is difficult, if not impossible, because certain information related to it such as the distances of the paths (that depend on the location of the cells and machines on the ground of the workshop) and the manufacturing times (influenced by the scheduling decisions) are unknown. The authors then end up using the similarity coefficients as an indirect measure of efficiency.

\subsection{Comments on the criteria}
\begin{figure*}
\centering\includegraphics[scale=0.8]{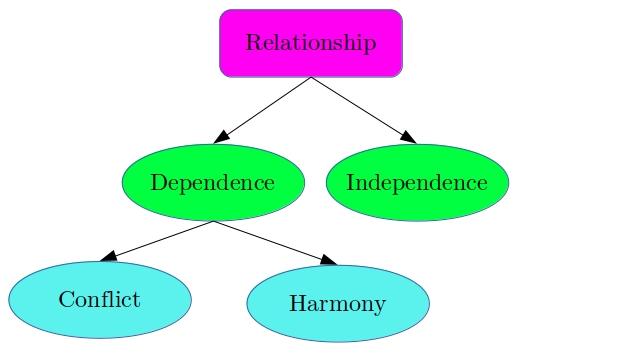} 
\caption{Criterion relationships.}
\label{critRelation}
\end{figure*}

After having gone over the criteria proposed in the literature, someone would be tempted to develop a model including this plethora of criteria. However, considering a large number of criteria can lead to solving difficulties. In particular, when it conforms to the Pareto optimisation spirit, because the size of the Pareto front becomes too large as the number of criteria increases [Fonseca1998, Deb2001].\\
If we are led to operate a reduction in the number of objectives, it is necessary to undertake it with a rigorous approach. Among the efforts deployed in this context, there is the work carried out in [Purshouse2003]. The author tries to bring a procedural approach to what a researcher confronted with a problem with a consistent number of criteria does in a non-formal way. It classifies the relationships that can link the criteria into three categories: independence, harmony and conflict (see Figure \ref{critRelation} \cite{Purshouse2003}). In the light of these relationships, the criteria are analysed in order to keep only a limited subset. Two criteria are in conflict if the improvement of one causes the degradation of the other. They are in harmony if the improvement of one leads to the improvement of the other. Finally, independence describes the fact that an improvement on one of the criteria has no influence on the other.\\ When it comes to materialisation, the author exploits these different relationships by relying on a sample taken from the population of solutions provided by his evolutionary solving approach. However, such an approach can lead to biased deductions, since the sample may not be representative. Furthermore, what is it that prevents the conflicting or harmonic features of two given criteria from depending on the instances of the problem: i.e. for some there is conflict, for others there is harmony?

Another approach we propose is based on studying redundancy and inconsistency that can be present when we compare some of the criteria.\\
For our problem, the use of an indirect definition for the efficiency criterion makes it redundant, since with the previous definition, it becomes equivalent to the similarity criterion (see paragraph \ref{maxFlex}). In addition, the criterion of minimisation of special elements is redundant because it is taken into account by the criterion of intercellular movements. The latter is in turn redundant, because it is taken into account by the intercellular traffic criterion.\\
On the other hand, we can identify a case of inconsistency in the simultaneous consideration of the criteria for minimising inter and intracellular movements. Indeed, the movements of the products can be deduced directly from the routings. As long as a movement is either intercellular or intracellular, by considering fixed routings during the entire planning horizon, the sum of these movements is always constant. Some researchers [Gupta1995] have opted for the combination of these two criteria in a multi-criteria model using the weighted sum technique. We can argue that actually, this combination allows only one criterion to be optimised. Indeed, the sum of the inter and intracellular movements being constant, assuming $f_1$ and $f_2$ the objective functions measuring the inter and intracellular movements respectively, we will have:\\
$Min\ Z=w_1\cdot f_1+w_2\cdot f_2 \Leftrightarrow Min\ Z= w_1\cdot f_1+w_2\cdot (c-f_1)$\\  
This is derived from putting $c= f_1+f_2$. Hence,\\ 
$Min\ Z=w_1\cdot f_1+w_2\cdot f_2\ \Leftrightarrow Min\ Z=(w_1-w_2)\cdot f_1+w_2\cdot c$
${  }\ {  }{  }\ {  }\ {  }\ {  }\ {  }\ {  }\ {  }\ {  }\ {  }\ {  }\ {  }\ {  }\ {  }\ {  }\ {  }\ {  }\ {  }\ {  }\ {  }\ {  }\ {  }\ {  }\ {  }\ {  }\ {  }\ {  }\ {  }\ {  }\ {  }\ {  }\ \Leftrightarrow Min\ Z'=(w_1-w_2) \cdot f_1$\\
because the term $w_2 \cdot c$ is constant.
However,$w_1$ is always set to be greater than $w_2$ because it is commonly accepted that minimising intercellular traffic is by far more important than minimising intracellular traffic. Hence,\\
 
\begin{equation}
\resizebox{0.4\textwidth}{!}{
$Min\ Z = w_1 \cdot  f_1 + w_2 \cdot  f_2 \Leftrightarrow Min\ Z'' = f_1$}
\label{intraInterCellAggreg}
\end{equation}

This development therefore permits to realise that, contrary to first appearances, minimising the weighted sum of these criteria is in fact minimising only one of them. Note that the papers \cite{Gupta1995, Logendran1995} only consider the two mentioned criteria (The author of the overview devoted to multi-criteria approaches \cite{Mansouri2000} nevertheless included the reference \cite{Gupta1995} in his work).\\
The most widely used reduction approach is may be the one that attempts to analyse the different combinations of objectives adopted in the literature. In \cite{Mansouri2000}, the author proposes to retain the following criteria: the cost of machine duplication, intercellular traffic, number of intercellular movements and intracellular workload imbalance. The survey we carried out, adding more than thirty articles, shows that among the three mentioned criteria, the minimisation of intercellular traffic, which is a generalisation of the criterion of the number of intercellular movements, is by far the most recurrent (see table \ref{critUtil}).

\begin{center}
\begin{table*}
\label{critUtil}
\caption{Utilisation of the criteria in published works}
\rowcolors{2}{white}{gray!25}
\centering
\begin{tabular}{c  p{0.03cm}  c  c  c  c  c  c  c  c  c  c  c  c  c  c  c}
\hline
\multicolumn{1}{c}{ } & & \multicolumn{15}{c}{Criteria} \\[2mm]
\cline{3-17}
Papers  & &  \rotatebox[origin=c]{70}{Special elem.} &  
        \rotatebox[origin=c]{70}{Intercel. mouv.} &
        \rotatebox[origin=c]{70}{Intracel. mouv.} &  
        \rotatebox[origin=c]{70}{Mach. duplic.} &  
        \rotatebox[origin=c]{70}{Subcontr. cost.} &  
        \rotatebox[origin=c]{70}{Intercel. traf} &  
        \rotatebox[origin=c]{70}{Mach. utilisat.} &  
        \rotatebox[origin=c]{70}{Manuf. cost} &  
        \rotatebox[origin=c]{70}{Intra. imbal.} &  
        \rotatebox[origin=c]{70}{Inter. imbal.} &
        \rotatebox[origin=c]{70}{Intracel. avoid.} &  
        \rotatebox[origin=c]{70}{Prouct Simil.} &  
        \rotatebox[origin=c]{70}{Reconfig. cost.} &  
        \rotatebox[origin=c]{70}{Flexibility} &  
        \rotatebox[origin=c]{70}{Efficacy} \\
\cline{1-1}\cline{3-17}
{\cite{Adil1996}} & & { *} &   &   &   &   &   &   &   &   &   &   &   &   &   &  \\
{\cite{Akturk1996}} & &   & { *} &   &   &   &   &   &   & { *} & { *} & { *} &   &   &   &  \\
{\cite{AlBadawi2005}} & &   &   &   &   &   &   &   &   &   &   &   & { *} &   &   &  \\
{ \cite{Al-Sultan1997}} & &   &   &   &   &   &   &   &   &   &   &   & { *} &   &   &  \\
{ \cite{Askin1990}} & &  &   &   & { *} &   & { *} &   &   &   &   &   &   &   &   &  \\
{ \cite{Atmani1995}} & &  &   &   &   &   & { *} &   & { *} &   &   &   &   &   &   &  \\
{ \cite{Boctor1991}} & &{ *} &   &   &   &   &   &   &   &   &   &   &   &   &   &  \\
{ \cite{Boulif2008}} &  & &   &   &   &   & { *} &   &   &   &   &   &   & { *} &   &  \\
{ \cite{Cao2004}} & &  &   &   & { *} &   & { *} &   &   &   &   &   &   &   &   &  \\
{ \cite{Chen1998}} & &  &   &   & { *} &   & { *} &   &   &   &   &   &   & { *} &   &  \\
{ \cite{Chen1999}} &  & &   &   &   &   & { *} & { *} &   &   &   &   &   &   &   &  \\
{ \cite{Chen1995}} & &  &   &   &   &   &   & { *} &   &   &   &   &   &   &   &  \\
{ \cite{Dahel1993}} & &  & { *} &   &   &   &   &   &   &   &   &   &   &   &   &  \\
{ \cite{Defersha2008}} & &   &   &   & { *} &   & { *} &   & { *} &   &   &   &   & { *} &   &  \\
{ \cite{Goncalves2002}} & &   & { *} &   &   &   &   & { *} &   &   &   &   &   &   &   &  \\
{ \cite{Gupta1995}} &  & & { *} & { *} &   &   &   &   &   &   &   &   &   &   &   &  \\
{ \cite{Gupta1996}} & &  &   & { *} &   &   &   &   &   & { *} &   &   &   &   &   &  \\
{ \cite{Harhalakis1990}} & &   &   &   &   &   & { *} &   &   &   &   &   &   &   &   &  \\
{ \cite{Harhalakis1994}} & &   &   &   &   &   & { *} &   &   &   &   &   &   &   &   &  \\
{ \cite{Ho1996}} & &  &   &   &   &   &   &   & { *} &   &   &   &   &   &   &  \\
{ \cite{Irani1993}} & &   & { *} &   &   &   & { *} &   &   &   &   &   &   &   &   &  \\
{ \cite{Jaya1998}} & &   &   &   &   &   &   &   &   &   &   &   & { *} &   &   &  \\
{ \cite{Joines1997}} & &  &   &   &   &   &   & { *} &   &   &   &   &   &   &   &  \\
{ \cite{Lee1997}} & &  &   &   &   &   &   &   &   &   & { *} &   &   &   &   &  \\
{ \cite{Liang1995}} & &   &   &   &   &   &   &   &   &   &   &   & { *} &   & { *} & { *}\\
{ \cite{Logendran1991}} & &   &   & { *} &   &   &   &   &   &   &   &   &   &   &   &  \\
{ \cite{Logendran1993}} & &   & { *} & { *} &   &   &   & { *} &   &   &   &   &   &   &   &  \\
{ \cite{Logendran1995}} & &   & { *} & { *} &   &   &   &   &   &   &   &   &   &   &   &  \\
{ \cite{Logendran1997}} & &  &   &   & { *} & { *} &   &   &   &   &   &   &   &   &   &  \\
{ \cite{Mansouri2002}} & &  & { *} &   & { *} &   &   & { *} &   &   & { *} &   &   &   &   &  \\
{ \cite{Min1993}} & &  &   &   &   &   &   &   & { *} &   &   &   & { *} &   &   &  \\
{ \cite{Mukhopadhyay1995}} & &  &   &   &   &   &   & { *} &   &   &   &   &   &   &   &  \\
{ \cite{Murthy1995}} & & { *} &   &   &   &   &   &   &   &   &   &   &   &   &   &  \\
{ \cite{Moon1999}} & &  &   &   & { *} &   &   &   & { *} &   &   &   &   &   &   &  \\
{ \cite{Nagi1990}} & &  &   &   &   &   & { *} &   &   &   &   &   &   &   &   &  \\
{ \cite{Pierreval1998}} & &   &   &   &   &   & { *} &   &   &   & { *} &   &   &   &   &  \\
{ \cite{Rajagopalan1975}} & &   & { *} &   &   &   &   & { *} &   & { *} &   &   &   &   &   &  \\
{ \cite{Rajamani1996}} & &  &   &   &   &   &   &   & { *} &   &   &   & { *} &   &   &  \\
{ \cite{Sankaran1993}} &  & &   &   &   &   &   &   & { *} &   &   &   &   &   &   &  \\
{ \cite{Shafer1992}} & &  & { *} &   &   & { *} &   &   &   &   &   &   &   &   &   &  \\
{ \cite{Seifoddini1989}} &  &   &   &   &   &   &   &   &   &   &   &   & { *} &   &   &  \\
{ \cite{Sofianopoulou1997}} & &   & { *} &   &   &   &   &   &   &   &   &   &   &   &   &  \\
{ \cite{Su1998}} & &  &   &   & { *} &   & { *} &   &   & { *} & { *} &   &   &   &   &  \\
{ \cite{Sureh1995}} & &   & { *} &   & { *} &   &   &   &   &   &   &   &   &   &   &  \\
{ \cite{Taboun1991}} & &   &   &   &   &   &   &   &   &   &   &   & { *} &   &   &  \\
{ \cite{Taboun1998}} & &  &   &   &   & { *} & { *} &   &   &   &   &   &   &   &   &  \\
{ \cite{Vakharia1990}} & &   &   &   &   &   &   &   &   &   &   &   & { *} &   &   &  \\
{ \cite{Venugopal1992a}} & &   &   &   &   &   & { *} &   &   & { *} &   &   &   &   &   &  \\
{ \cite{Venugopal1992b}} & &  &   &   &   &   &  &   &   & { *} &   &   &   &   &   &  \\
{ \cite{Verma1995}} &  &  & { *} & { *} &   &   &   &   &   &   &   & { *} &   &   &   &  \\
{ \cite{Vin2003}} &  &  &   &   &   &   &   & { *} &   &   &   &   & { *} &   & { *} &  \\
{ \cite{Wei1990}} & &  &   &   &   & { *} &   &   &   & { *} & { *} &   &   &   &   &  \\
{ \cite{Wicks1999}} & &   &   &   & { *} &   & { *} &   &   &   &   &   &   & { *} &   &  \\
{ \cite{Zhao2000}} & &  & { *} & { *} &   &   &   &   &   & { *} &   &   &   &   &   &  \\
{ \cite{Zolfaghari1997}} &  &  &   &   &   &   &   & { *} &   &   &   &   &   &   &   &  \\
\hline
\multicolumn{2}{c}{Utilis.(\%)} & 05.45 & \bf{25.45} & 12.73 & 18.18 & 07.27 & \bf{29.09} & 18.18 & 12.73 & 14.55 & 10.91 & 03.64 & 18.18 & 07.27 & 03.64 & 01.82 \\
\hline
\end{tabular}
\end{table*}
\end{center}

\begin{figure*}
\centering\includegraphics[scale=0.9]{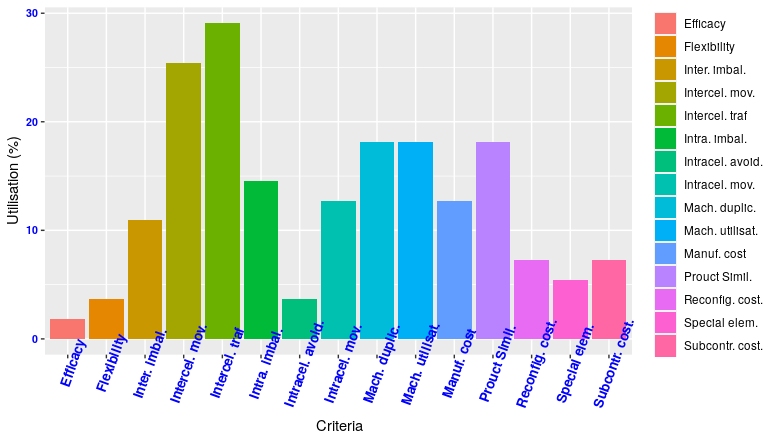} 
\caption{Criterion percentage of presence in surveyed research works.}
\label{critUtilFig}
\end{figure*}

The analysis of the criteria allows us to retain the following criteria consisting of minimising:
\begin{enumerate}
    \item intercellular traffic;
    \item intercellular workload imbalance;
    \item intracellular workload imbalance;
and to add to them if applicable:
    \item manufacturing cost of products (routing flexibility approaches);
    \item the reconfiguration cost (dynamic approaches);
\end{enumerate}
This choice is justified by the following reasons:
\begin{itemize}
    \item The criterion of minimising intercellular traffic is taken into consideration in most of the surveyed papers (see Fig. \ref{critUtilFig}).
    \item minimisation of workload imbalance is a criterion intended to attenuate the side effects of the minimisation of traffic on the balancing of intra and intercell workload.
    \item The criteria for minimising special elements and intercellular movements can be omitted because the criterion of intercellular traffic is a generalisation that compensates for their shortcomings.
    \item The criterion of minimisation of intracellular movements was omitted for impossibility of conciliation with the main criterion (see formula \ref{intraInterCellAggreg} consequence).
    \item Duplication of machines may not be considered as it is not always plausible due to the high cost of machines. This has the effect of weakening the contribution of certain criteria which have been added only to compensate for the side effects of duplication, namely the minimisation of the costs of duplication, avoidance and under-use of machines. Indeed, it is the duplication which generates problems of circumvention of machines inside the cells [Akturk1996]. It is also at the origin of the under-utilisation of machines due to the excess of capacity.
\end{itemize}

Finally, note that the reduction of criteria remains an open problem for which we propose an approach which consists in trying to develop relationships by relying on an analytical representation of the criteria, in relation to each other. This representation will make it possible to detect the type of relationship linking the different criteria (conflicting, harmonic or not reconcilable). Such a representation becomes more and more difficult when the criteria do not use the same inputs. In this case, a study of correlation can be considered.

\section{Conclusion}
In order to better understand the problem of cell composition, we have identified the different decisions to be taken for its resolution, the practical constraints to be considered and the criteria not to be overlooked. Throughout the overview, some open issues were exposed, such as the development of an integrated approach including all decisions in a dynamic way and the adoption of a scientifically based criteria reduction approach. So many questions to be developed which inform that the problem of cell composition is not yet ready to reveal all its secrets.   
   


\end{document}